\documentstyle[12pt]{article}
\def\CA{{\cal A}}   \def\CD{{\cal D}} 
 \def\CF{{\cal F}}   
   \def\CL{{\cal L}} 
 \def\CN{{\cal N}} \def\CO{{\cal O}} \def\CP{{\cal P}}

\def\BA{{\bf A}} \def\BB{{\bf B}}  \def\BD{{\bf D}} 
    
 \def\BJ{{\bf J}}  \def\BL{{\bf L}} 
\def\BM{{\bf M}} \def\BN{{\bf N}}  \def\BP{{\bf P}}
   
   \def\BX{{\bf X}}
 
\def\Ba{{\bf a}}

   \def\Bx{{\bf x}}
 
\newfont{\frak}{eufm10 scaled 1200}
\newcommand{\mfrak}[1]{\mbox{\frak #1}}

\def\half{{1\over2}}
\def\qed{\hfill{\sl q.e.d}\medskip}
\def\Tr#1{{\rm T}\!{\rm r}\{#1\}}
\def\id{{\bf1}}
\def\mat#1{\left(\matrix{#1}\right)}
\def\proof{\medskip\noindent{\sl Proof\ } :\  }
\def\norm#1{|\!|#1 |\!|}

\def\complex{{\mathchoice
{\setbox0=\hbox{$\displaystyle\rm C$}\hbox{\hbox to0pt
{\kern0.4\wd0\vrule height0.9\ht0\hss}\box0}}
{\setbox0=\hbox{$\textstyle\rm C$}\hbox{\hbox to0pt
{\kern0.4\wd0\vrule height0.9\ht0\hss}\box0}}
{\setbox0=\hbox{$\scriptstyle\rm C$}\hbox{\hbox to0pt
{\kern0.4\wd0\vrule height0.9\ht0\hss}\box0}}
{\setbox0=\hbox{$\scriptscriptstyle\rm C$}\hbox{\hbox to0pt
{\kern0.4\wd0\vrule height0.9\ht0\hss}\box0}}}}
\def\Co{{\mathchoice
{\setbox0=\hbox{$\displaystyle\rm C$}\hbox{\hbox to0pt
{\kern0.4\wd0\vrule height0.9\ht0\hss}\box0}}
{\setbox0=\hbox{$\textstyle\rm C$}\hbox{\hbox to0pt
{\kern0.4\wd0\vrule height0.9\ht0\hss}\box0}}
{\setbox0=\hbox{$\scriptstyle\rm C$}\hbox{\hbox to0pt
{\kern0.4\wd0\vrule height0.9\ht0\hss}\box0}}
{\setbox0=\hbox{$\scriptscriptstyle\rm C$}\hbox{\hbox to0pt
{\kern0.4\wd0\vrule height0.9\ht0\hss}\box0}}}}
\def\Rl{{\mathchoice
{\setbox0=\hbox{$\displaystyle\rm R$}\hbox{\hbox to0pt
{\kern0.4\wd0\vrule height0.9\ht0\hss}\box0}}
{\setbox0=\hbox{$\textstyle\rm R$}\hbox{\hbox to0pt
{\kern0.4\wd0\vrule height0.9\ht0\hss}\box0}}
{\setbox0=\hbox{$\scriptstyle\rm R$}\hbox{\hbox to0pt
{\kern0.4\wd0\vrule height0.9\ht0\hss}\box0}}
{\setbox0=\hbox{$\scriptscriptstyle\rm R$}\hbox{\hbox to0pt
{\kern0.4\wd0\vrule height0.9\ht0\hss}\box0}}}}

\def\be{\begin{equation}}
\def\ee{\end{equation}}
\def\bea{\begin{eqnarray}}
\def\eea{\end{eqnarray}}

\def\hilbert{\mfrak{H}}

\begin{document} 

\begin{titlepage}
\vspace*{-4ex}

\null \hfill Preprint TU-498  \\
\null \hfill March 1996 \\[4ex]

\begin{center}

\bigskip
\bigskip

{\Large \bf
Chirality and Dirac Operator on \\[1.5em]
Noncommutative Sphere
}\\[3ex]

Ursula Carow-Watamura\footnote{Fellow of the Japan Society for Promotion of 
Science, e-mail: ursula@tuhep.phys.tohoku.ac.jp} \  and \  
Satoshi Watamura\footnote{e-mail: watamura@tuhep.phys.tohoku.ac.jp}

Department of Physics \\
Graduate School of Science \\
Tohoku University \\
Aoba-ku, Sendai 980-77, JAPAN \\ [2ex]
\end{center}

\bigskip
\begin{abstract}
\medskip
We give a derivation of the Dirac operator on the 
noncommutative $2$-sphere within the framework of the 
bosonic fuzzy sphere and define Connes' triple. It turns out that there 
are two different types of 
spectra of the Dirac operator and correspondingly there are two 
classes of quantized algebras. As a result we obtain a new 
restriction on the Planck constant in Berezin's quantization.
The map to the local frame in noncommutative geometry
 is also discussed.

\end{abstract}
\end{titlepage}
\eject  


\section{Introduction}

The description of spacetime at the order of the Planck scale and 
the description of the nature of quantum gravity is a long-standing 
problem. Quite a number of proposals have been made in order to 
describe consistently gravitational interaction and quantum field 
theory, however these proposals either do not give a satisfying formulation 
of the quantum theory of gravity or, in their present form an 
interpretation as a theory of the geometry is difficult.

Thus, recently the 
modification of the concept of geometry itself is also discussed
by many authors.
Of course one may argue that a successful theory of gravitation will
naturally exhibit the necessary structure of a suitable theory of 
geometry and a natural language to describe it. 
On the other hand it is not very probable that the standard language of 
ordinary differential geometry is a suitable tool. 
The noncommutative geometry from 
physicist's point of view is a possibility 
to describe such a geometry.

The noncommutative geometry is discussed in many contexts. 
The common idea is that one deals with a function algebra over 
the space one is interested in and the description 
of its geometry is made free from 
the concept of a point \cite{ConnesNCG}. 
In other words, 
the geometry of a manifold is reformulated in terms of an algebra of 
functions defined over it which may be called 
structure algebra.
Once the algebraic description of the 
geometry is obtained, the structure algebra can be made 
noncommutative.

As an application for physics, recently Connes proposed 
an interpretation of the standard model within the framework 
of noncommutative geometry \cite{ConnLott90}.  
The Higgs field in the standard model 
is interpreted as a gauge field in the noncommutative
space given by a product of Minkowski space and a two-point space.  
This means that, in an appropriate geometry there is a chance 
to understand all bose particles through the gauge principle 
\cite{ChamFeld93}. 
On the other hand we may consider this formulation of the standard model 
as a kind of Kaluza-Klein theory with the noncommutative space 
playing the role of an internal space. 
This idea is also natural since the scale of the internal space is the 
Planck scale where we expect the breakdown of the classical picture of 
geometry. (See for example \cite{MadoMour96} 
and references therein.) 
From this point of view  
it is very interesting 
to investigate the other possibilities 
where we consider as an internal space a more complicated 
object such as the noncommutative sphere. 
With the above problem in mind, in the present paper we try to find a 
proper formulation of the quantum sphere which fits to 
Connes' idea. 

When discussing the noncommutative analogue of the 
sphere, so far there are two methods 
available. One is the q-deformation of the sphere 
\cite{Pod} and the other one, which is the case considered here, 
is the Berezin (or Berezin-Toeplitz) quantization of the sphere 
\cite{Berezin74}, which is recently also known as the fuzzy 
sphere \cite{MadoreCQG92}.

In order to define the differential calculus within 
Connes' framework one has to construct the triple 
$(\CA, \CD, \hilbert)$, where $(\hilbert,\CD)$ is a $K$-cycle over 
the algebra $\CA$ \cite{ConnesNCG}. 
$\hilbert$ is a Hilbert space, $\CA$ is an 
algebra of bounded operators acting on this Hilbert space, 
and $\CD$ is a linear operator on $\hilbert$, the Dirac operator. 
The Hilbert space is graded (i.e. we consider the even case).  
The algebra $\CA$ is even with respect to the 
grading and the Dirac operator is odd.
Once this $K$-cycle is given, the construction of the differential calculus 
is rather straightforward.

The aim of this paper is to define 
a Dirac operator as well as a chirality operator on the 
noncommutative sphere such that it can be 
naturally combined with the 
field theory in Connes' framework. 

The Dirac operator on the fuzzy sphere has been also discussed in 
refs.\cite{GrosMado92,GrosMado95,GrosKlim95}, where 
the supersymmetric extension of the fuzzy sphere is constructed 
using the supersymmetry algebra $OSp(2,1)$; then the Dirac operator is
 defined on that super fuzzy sphere. 
On the contrary, our construction of the 
Dirac operator does not make use of the supersymmetry algebra, i.e., 
we use the algebra of the bosonic fuzzy sphere. 
The resulting chirality operator, the Dirac operator and 
the definition of the spinors are different 
from the ones obtained by the supersymmetry algebra approach. 

Our approach is the following: We define the algebra $\CA_N$ 
of the fuzzy sphere 
which is generated by noncommutative operators $\Bx_i$, ($i=1,2,3$) 
where $N$ is an integer relating to the Planck constant 
in Berezin's quantization. 
Then we introduce the Lie algebra $\CL_N$ defined by 
the derivations $\BL_i$ given by the adjoint action of the generators.
Including the derivations we consider the bigger algebra $(\CA_N,\CL_N)$ 
generated by $(\Bx_i,\BL_i)$.  The chirality operator and 
the Dirac operator are constructed
algebraically in the algebra $(\CA_N,\CL_N)\otimes M_2(\complex)$ 
where $M_2(\complex)$ is the algebra of complex $2\times2$ matrices. 

Then using the result, we investigate its properties such as the spectrum. 
We reconsider the algebra of the noncommutative sphere and 
obtain the corresponding triple which we need for applying Connes' approach. 
In the naive commutative limit, 
the structure of the spectrum reveals the
fine structure of Berezin's quantization and gives the restriction on
the quantization parameter $N$ to be even integer. 
Furthermore, we also discuss the Dirac operator in the local patch 
within the framework of noncommutative geometry.

The organization of this paper is as follows. In section 2, we give a 
brief review of the operator algebra of the fuzzy sphere and introduce 
notations. In section 3 we derive chirality operator and Dirac operator 
and examine its properties and the Hilbert space structure. 
Section 4 is devoted to discussions and conclusions.

\section{Algebra of Fuzzy Sphere and Derivations}

\subsection{Brief Summary on Fuzzy Sphere}

The noncommutative sphere has  been considered by several authors in 
different contexts such as an example for a general quantization 
procedure \cite{Berezin74,Berezin75a} (see also for example 
\cite{CaheGutt90,BordHopp91,Coburn92,KlimLes92,BordMein94} 
 and references therein), 
the algebra appearing in membrane \cite{Hoppe89}, 
in relation with coherent states \cite{Perelomov72}, and recently 
in connection with noncommutative geometry 
\cite{MadoreCQG92,GrosMado95,GrosKlim95}. The same structure also appears 
in the context of the quantum Hall effect \cite{Haldane83,FanoOrto86}. 
The noncommutative sphere is described in various ways, however 
the resulting algebra is the same. 
It is easy to understand the idea of the fuzzy sphere from the point 
of view of approximation. 

It is well known that 
the square integrable functions on a $2$-sphere $\CL^2(S^2)$ 
form a Hilbert space the basis of which is given by the 
spherical harmonics $Y_{lm}$.
By usual multiplication these functions form a closed algebra. Thus 
we have a basic function algebra $\CA_\infty$ over the sphere, and
any element of the algebra can be expanded in the basis 
of spherical harmonics.
The idea of the fuzzy sphere may be formulated in brief as that 
we approximate the functions
on the sphere by a finite number of spherical harmonics 
where this number is limited by the 
maximal angular momentum $\{Y_{lm}; l\leq N\}$. 
However with respect to the usual multiplication 
this set of functions does not form a closed algebra 
since the product of two spherical harmonics $Y_{lm}$
and $Y_{l'm'}$ contains $Y_{l+l',m}$ due to the product rule.  
It is a new multiplication rule that cures the above 
described situation and 
gives a 
closed function algebra with a finite number of basis elements.
The resulting algebra $\CA_N$ is noncommutative.\footnote{We denote
the algebra of fuzzy sphere as 
$\CA_N$ with suffix $N$ which becomes the eigenvalue
of the number operator defined in the next section.}
The interpretation of this property is that we obtain a geometry 
where the concept of a point is "dissolved".\footnote{The "fuzziness" 
of the sphere is removed by taking the limit $N\rightarrow\infty$, in 
which the function algebra becomes commutative.} 
Since the number of the independent functions with angular momentum
$l\leq N$ is $(N+1)^2$, as a vector space this has the same dimension 
as the vector space spanned by the 
$(N+1)\times(N+1)$ matrices. This is not accidental and we 
can identify
the algebra of the fuzzy sphere with the algebra of complex matrices 
$M_{N+1}(\complex)$ and thus we can consider it as a special case of the 
matrix geometry\cite{DuboKern89a,DuboKern90,DuboKern90a,DuboMado91}, 
which is the way how the fuzzy sphere is introduced 
in ref.\cite{MadoreCQG92}.

In the above description the truncation of the function algebra which leads 
to the noncommutative algebra is
rather ad hoc. However the construction of this type of noncommutative
algebra is equivalent to Berezin's quantization and can be generalized
to a general K\"ahler manifold.
In the Berezin-Toeplitz 
quantization one quantizes the algebra using the Poisson structure 
on the manifold defined 
by the K\"ahler form.  For this,
 one considers a finite dimensional Hilbert space $\CF_N=\Gamma(L^N)$, 
given by holomorphic sections on $L^N$, 
where $L^N=\otimes^N L$ is the $N$th tensor of the 
line bundle $L$.
Then by using the coherent states $|v\rangle$ where $(v,\overline{v})$
parametrize $S^2$ (see below),
one can define an operator acting on the Hilbert space $\CF_N$ 
for any function $f(v,\overline{v})\in\CL^2(S^2)$ by 

\be
T_f|v\rangle=\CP f|v\rangle\ ,
\ee
where $\CP$ is the projection operator from the general sections 
 to the subspace $\CF_N$, 
$$
\CP: \Gamma(L^{\otimes N})\rightarrow \CF_N \quad .
$$ 
The resulting operator is defined on $\CF_N$.  The product of two operators 
is defined by successive application of the above construction
\be
T_fT_g|v>=\CP f\CP g|v>.
\ee
and with this multiplication we define the algebra $\CA_N$ of the operators 
$T_f$. $T_f$ is called Toeplitz operator. From the above definition one 
easily sees that the multiplication of operators $T_f$ is in general 
noncommutative.  Furthermore, their commutator satisfies 
Berezin's quantization condition \cite{Berezin74}
\be
\lim_{N\rightarrow\infty}{1\over N}[T_f,T_g]=\{f,g\}_{PB} \ ,
\ee
where the r.h.s. is the Poisson bracket defined through the K\"ahler form
and thus the resulting algebra approximates the Poisson algebra of the 
function \cite{CaheGutt90,KlimLes92,BordMein94}. See also \cite{Madore91}. 

By taking an appropriate basis in the Hilbert space $\CF_N$ one can represent
the Toepliz operator by a matrix.  For the case of a 
$2$-sphere the dimension of 
$\CF_N$ is $N+1$ and thus we obtain an algebra 
of $(N+1)\times (N+1)$ matrices which is a representation of the
algebra discussed in ref.\cite{MadoreCQG92}.  On the other hand, 
by defining coherent states and representing the Toepliz operator
as an expectation value with respect to these coherent states 
${<z|T_f|z>\over<z|z>}$ which is 
called a covariant symbol \cite{Berezin72}, we obtain Berezin's quantized 
algebra. (See also \cite{Perelomov72,GrosPres93}.)  
The product of this algebra, i.e. the $*$-product among the
covariant symbols is simply defined by rewriting 
the above operator multiplication in the language of the covariant symbols 
and
thus it is in general noncommutative as is the multiplication of 
operators. In this way, 
the algebra of the fuzzy sphere can be understood as a finite approximation of 
the function algebra. The problem is to clarify within which framework 
the differential calculus must be introduced.

\subsection{Operator Representation of the Algebra and its Derivations}

As we discussed above, when considering the algebra of 
the noncommutative sphere,
it is natural to start with the algebra of
 operators acting on the Hilbert space
$\CF_N$. Since $\CF_N$ is a representation space of the rotation group, 
we introduce in the standard way a pair of creation-annihilation
 operators $\Ba^{\dagger b}, \Ba_b$ ($b=1,2$) 
which transform as a fundamental representation under the $SU(2)$ action, 
satisfying 

\be
[\Ba^a,\Ba^\dagger_b]=\delta^a_b \ .\label{BosonicCommutator}
\ee  
Define the number operator by
\be
\BN=\Ba^\dagger_b\Ba^b \ ,
\ee
then a set of states satisfying
\be
\BN|v>=N|v>
\ee
provides an $N+1$ dimensional irreducible representation space $\CF_N$.
The operator algebra $\CA_N$ on $\CF_N$ 
is unital and given by the operators $\{\CO; [\BN,\CO]=0\}$.
The  generators of the algebra $\CA_N$ are defined by 
\be
\Bx_i=\half\alpha\sigma_i^{a}{}_b\Ba^\dagger_a\Ba^b \ ,
\ee
where the normalization factor $\alpha$ 
is a central element $[\alpha, \Bx_i]=0$ introduced for later 
convenience \footnote{Formally we have to include $\BN$ and $\alpha$ 
as generators of the algebra, however they can be treated as 
numbers in the following calculations.}. 
The commutation relations among these operators are 
\bea
[\Bx_i,\Ba^\dagger_a]&=&\half\alpha \sigma_i^{b}{}_{a}\Ba^\dagger_b\ ,\cr
[\Bx_i,\Ba^a]&=&-\half\alpha \sigma_i^{a}{}_b \Ba^b\ .
\eea
The algebra of the fuzzy sphere is generated by $\Bx^i$ and the basic
relation is
\be
[\Bx_i, \Bx_j]=i\alpha\epsilon_{ijk}\Bx_k.\label{fuzzyalgebra}
\ee
The normalization $\alpha$ is defined by 
\be
\Bx_i\Bx_i={\alpha^2\over4}\BN(\BN+2)=\ell^2.\label{radius}
\ee
This means that $\ell>0$ is the radius of the $2$-sphere and we get
\be
\alpha={2\ell\over\sqrt{\BN(\BN+2)}}\ .\label{alpha}
\ee

Note that in the formulation on algebra level we use the number operator 
$\BN$ in order to keep 
the independence of the algebra from the representation space $\CF_N$.
Thus the "Planck constant" $\alpha$ is also an operator.  However, 
when we discuss the commutative limit $N\rightarrow\infty$, we take a definite 
Hilbert space $\CF_N$ and thus the number operator is replaced by 
its eigenvalue $N$. The Planck constant $\alpha$ becomes also
a number ${2\ell\over\sqrt{N(N+2)}}$. Then, we take the limit 
$N\rightarrow\infty$, i.e., $\alpha\rightarrow0$. 
Therefore, the commutative limit discussed in the later part of 
this paper 
is taken with 
respect to the eigenvalue $N$ of the number operator on the space $\CF_N$.

\noindent{\sl Proof} of eq.(\ref{radius}):

From the Fierz identity we get
\bea
\BN^2+{4\over\alpha^2}\Bx_i\Bx_i&=&\Ba^\dagger_a(\sigma_\mu)^a{}_b\Ba^b
\Ba^\dagger_d(\sigma_\mu)^d{}_c\Ba^c\cr
&=&2\Ba^\dagger_a\Ba^b\Ba^\dagger_d\Ba^c
\delta^a_c\delta^d_b\cr
&=&2\Ba^\dagger_a(\BN+2)\Ba^c
\delta^a_c=2(\BN+1)\BN\ ,
\eea
where $\mu=0,1,2,3$ and $\sigma_0^a{}_b=\delta^a_b$. We have also used
\be
\Ba^a\Ba^\dagger_a=\delta^a_a+\Ba^\dagger_a\Ba^a=2+\BN \ ,
\ee
\be
\BN\Ba^\dagger=\Ba^\dagger(\BN+1) \ .
\ee
Thus we get the above relation.
\qed

Now let us consider the derivations of 
$\CA_N$. 
The derivations can be defined by the commutator with 
any algebra element in $\CA_N$ since the adjoint action always 
defines an inner derivation.  
We introduce the derivative operator $\BL_i$ by the adjoint action 
of $\Bx^i$ \cite{MadoreCQG92}
  
\be
{1\over \alpha}ad_{\Bx_i}\Bx_k
={1\over \alpha}[\Bx_i,\Bx_k] \equiv \BL_i \Bx_k\ .
\ee
These objects are the noncommutative analogue of the
Killing vector fields on the sphere and the  algebra of $\BL_i$ closes 
and we obtain a Lie algebra $\CL_N\subset Der(\CA_N)$ 
(see also ref.\cite{DuboKern90}).

Since the action of $\BL_i$ generates rotations 
of the noncommutative sphere, 
$\BL_i$ and $\Bx_k$ satisfy
the usual commutation relations of $SU(2)$. 
Thus the algebra $(\CA_N,\CL_N)$ of the noncommutative sphere together 
with derivations
is defined by the operator 
algebra given by $\Bx_i$ and $\BL_i$  
\bea
[\BL_i,\Bx_j]=i\epsilon_{ijk}\Bx_k\ ,\qquad
[\BL_i,\BL_j]=i\epsilon_{ijk}\BL_k\ ,
\eea
together with (\ref{fuzzyalgebra}).

\subsection{Integration and Coherent State Representation}

The integration on the noncommutative sphere is defined as a trace over
the Hilbert space $\CF_N$.  The simplest way to take the 
trace is to use the orthogonal basis:
\be
|k;N\rangle ={1\over \sqrt{k!(N-k)!}}(\Ba^{\dagger}_1)^k
(\Ba^{\dagger}_2)^{N-k}|0\rangle\ ,
\label{Fockbasis} 
\ee
where $k=0,...,N$ and $|0\rangle$ is the vacuum.
Then 
\be
\langle\CO\rangle_N={1\over N+1}\Tr{\CO}
={1\over N+1}\sum_k \langle k;N|\CO|k;N\rangle \ .
\ee

In order to see the relation of the above result with 
integration in the commutative limit, 
we define the trace by using coherent states 
\cite{Bargmann61,Perelomov72,GrosPres93}.
The coherent states are parametrized by a point on the sphere. 
In order to keep track with the complex structure on the sphere, we take here
the parametrization obtained by projecting 
stereographically to the complex plane and
representing the corresponding point by complex 
coordinates $(z,\overline{z})$. Then
the coherent states are defined by

\be
\big|z;N\rangle={1\over\sqrt{N!}}
(z\Ba^\dagger_1+\Ba_2^\dagger)^{N}|0\rangle\ ,
\ee
which satisfies
\be
\langle  \overline{z};N|z;N\rangle 
=(1+\overline{z}z)^N \ ,
\ee
where we have taken the notation 
$|w\rangle^*=\langle w^*|=\langle \overline{w}|$.
Note that in the following we simply denote $|z;N\rangle$ as $|z\rangle$
for the coherent state in $\CF_N$.

Using the above definitions we can take the coherent state representation.
Then the orthogonal basis of $\CF_N$ can be represented 
by $z$:
\be
<z|k>=f_k(z)=\sqrt{N!\over k!(N-k)!} z^k\ .
\ee
The inner product of two state $|f\rangle$ and $|g\rangle$ is 
defined by
\be
 \langle\overline{f}|g\rangle
=\int d\mu_N(z) \overline{f(z)}g(z)\ ,
\ee
where the measure $\mu_N$ is defined such that it gives 
the orthonormality of the basis:
\be
\int d\mu_N(z)\overline{f_l(z)}f_k(z)
={(N+1)\over 2\pi i}\int dz\wedge d\overline{z} 
{f_k(z) \overline{f_l(z)}\over(1+|z|^2)^{N+2}}
=\delta_{kl}\ .
\ee

The reproducing kernel is given by
\be
L_N(z,\overline{v})=\sum f_k(z)\overline{f_k(v)}=\sum {N!\over k!(N-k)!}(z\overline{v})^k
=(1+z\overline{v})^N \ .
\ee

Thus in the coherent state representation, 
the trace over the Hilbert space $\CF_N$ is represented by
the integration of the symbol of an operator over $S^2$. 
\be
\langle\CO\rangle_N={1\over N+1}\Tr{\CO}=
{1\over N+1}\int d\mu_N <\overline{z}|\CO|z>\ .
\ee


\bigskip

\section{The Dirac operator}

\bigskip

\subsection{Chirality and Dirac Operator}

In this section we examine the algebraic 
relations among the operators to define the chirality operator 
and to construct 
the Dirac operator algebraically on the noncommutative sphere. 
For this purpose we consider the product algebra of $(\CA_N,\CL_N)$ and 
$\CA_1$. $\CA_1$ is simply the algebra of $2\times 2$ matrices $M_2(\complex)$
the elements of which can be represented by 
\be
\BM=\sum_{\mu=0}^4 a_\mu\sigma_\mu \ ,
\ee
and transform under rotation as
\be
\BM\rightarrow U\BM U^\dagger \ ,
\ee
where $U$ is the spin representation matrix of $SU(2)$.
Thus we define the chirality operator $\gamma_\chi$ and the Dirac operator
$\BD$ in the product algebra of $(\CA_N,\CL_N)\otimes M_2(\complex)$, i.e. 
$2\times 2$ matrices the entries of which are polynomials in 
$(\Bx_i,\BL_j)$.

Our strategy taken here is to define first the chirality operator and, once 
this is achieved the Dirac operator is constructed by the requirement 
that it should anticommute with this chirality operator. 

Thus let us first discuss the possibilities for defining 
the chirality operator.  
The simplest choice would be to take $\id\otimes\sigma^3$. 
However, this choice breaks the $SU(2)$ symmetry. For our purpose 
it is better to keep the $SU(2)$ symmetry and thus we take 
a rotational invariant operator as the chirality. 

On the commutative sphere, as is disscussed in ref.\cite{Jayewardena88} 
a natural chirality operator is
\be
\gamma_\infty={1\over \ell}\sum_i x_i \otimes \sigma_i\ ,
\ee
where $x_i$ is the homogenious coordinate and $\ell$ is the radius of the
sphere
($\sum_ix_ix_i=\ell^2$). Thus $\gamma_\infty^2=1$.
On the fuzzy sphere the coordinate function is replaced by
the operator $\Bx_i$. However, if we replace $x_i$ by the operator
$\Bx_i$ in the above definition, the square of the resulting operator is 
not unity due to the noncommutativity of $\Bx_i$. 

In order to discuss the above situation in detail and also 
for later convenience let us introduce the following 
operators in $(\CA_N,\CL_N)\otimes M_2(\complex)$:

\bea
\chi&=&\sum_i(\Bx_i\otimes\sigma_i)\ ,\cr
\Lambda&=&\sum_i(\BL_i\otimes\sigma_i)\ ,\cr
\Sigma&=&-i\sum_{ijk}\epsilon_{ijk}(\Bx_i\BL_j\otimes\sigma_k)\ .
\label{InvariantOperators}
\eea
These are all $SU(2)$ invariant operators.

As we discussed above, among the operators given in 
eq.(\ref{InvariantOperators}) 
$\chi$ is a good candidate for the chirality operator.
Here, however, since the coordinates are not commutative, the square of the
operator $\chi$ is not unity.
Instead, this operator satisfies the relation

\bea
\chi\chi
&=&(\Bx\cdot\Bx)\otimes\id-\alpha \chi\label{chichi}\ .
\eea
where $(\BA\cdot\BB)=\sum_{i=1}^3\BA_i\BB_i$. 
Therefore from eq.(\ref{chichi}) we obtain 
\be
(\chi+\half \alpha)^2=\chi^2+\alpha\chi+{1\over4}\alpha^2
=(\Bx\cdot\Bx)+{1\over4}\alpha^2\ .
\ee
This suggests that the chirality operator is given 
by 
\be
\gamma_\chi={1\over\CN_N}(\chi+\half \alpha)\ ,\label{chirality}
\ee
where the normalization constant is determined by the requirement 
\be
\gamma_\chi^2=1\ .
\ee
This gives 
\be
\CN_N^2=(\Bx\cdot\Bx)+{1\over4}\alpha^2\ ,
\ee
and thus
\be
\CN_N={\alpha\over2}(\BN+1)=\ell\sqrt{(\BN+1)^2\over\BN(\BN+2)}\ ,
\label{Normalization}
\ee
where we imposed relation (\ref{radius}). Thus, we define the chirality 
operator by eq.(\ref{chirality}) with normalization (\ref{Normalization}).

Once the chirality operator is chosen, 
the construction of the Dirac operator is rather straightforward.
For this end we have to 
consider the relations among the above operators. The relevant 
relations are as follows. 

The square of the other operators given in eq.(\ref{InvariantOperators}) 
satisfies 
\be
\Lambda\Lambda=(\BL\cdot\BL)\otimes\id-\Lambda\ ,
\ee
\be
\Sigma\Sigma=\alpha(\Bx\cdot\BL)-(\Bx\cdot\Bx)(\BL\cdot\BL)
+(\Bx\cdot\BL)(\Bx\cdot\BL)
-\alpha \Sigma+\alpha(\Bx\cdot\BL)\Lambda-(\Bx\cdot\Bx)\Lambda\ ,
\ee
and their commutators and anticommutators are given by
\bea
\{\chi,\Lambda\}&=&2[(\Bx\cdot\BL)\otimes\id-\chi]\ ,\\[0pt]
 [\chi,\Lambda]&=&2[\chi-\Sigma]\ ,
\eea
\bea
\{\Sigma,\chi\}&=&2(\Bx\cdot\Bx)\otimes\id-\alpha[\chi+\Sigma]\ ,\\[0pt]
 [\Sigma,\chi]
&=&2(\Bx\cdot\Bx)(\Lambda+\id)-2\alpha(\Bx\cdot\BL)\otimes\id
-\{(\Bx\cdot\BL),\chi\}\ ,
\eea
\bea
\{\Sigma,\Lambda\}&=&2(\Bx\cdot\BL)\otimes\id+2\Sigma\ ,\\[0pt]
 [\Sigma,\Lambda]&=&2(\Lambda(\Bx\cdot\BL)-\chi(\BL\cdot\BL))\ .
\eea

In order to determine the Dirac operator we make use of the 
requirement that it
must anticommute with the chirality operator. It turns out 
that this requirment defines
the Dirac operator rather uniquely. Combining 
the above relations we find the following identities. 

\be
\{\chi-\Sigma,\Lambda\}=2(\Sigma-\chi)\ ,
\ee
\be
\{\chi-\Sigma,\chi\}=\alpha(\Sigma-\chi) \ .
\ee
From the latter one we obtain

\be
\{\chi-\Sigma,\chi+\half\alpha\}=0 \ .
\ee

Thus there are two independent 
operators which anticommute with
the chirality operator $\gamma_\chi$ 
\be
(\Sigma-\chi)\ ,
\ee 
and 
\be
\gamma_\chi(\Sigma-\chi)\ .
\ee
These two are the candidates for the Dirac operator.  
It turns out that the second operator 
has similarity with the commutative case and thus 
we define the Dirac operator as
\be
\BD={1\over \ell}\gamma_\chi(\Sigma-\chi)
={\ell\over\CN_N}\Big\{(\Lambda+\id)-{1\over 2\ell^2}\{\chi,(\Bx\cdot\BL)\}
-{\alpha\over \ell^2}(\Bx\cdot\BL)\Big\}\ ,\label{dirac}
\ee
where $\CN_N$ is the normalization given in eq.(\ref{Normalization}).
It satisfies the required condition 
\be
\{\gamma_\chi,\BD\}=0 \ .
\ee
Note that the Dirac operator in this paper is taken to be dimensionless.

The first term on the r.h.s. in eq.(\ref{dirac}) 
has the same form as the Dirac operator for the commutative case. 
The other terms in eq.(\ref{dirac})
can be interpreted as ``quantum corrections''.  
Since 
\be
\lim_{N\rightarrow \infty} \CN_N=\ell \ ,
\ee 
in the naive 
limit the above Dirac operator has the standard form of the commutative 
case if the correction terms which include $(\Bx\cdot\BL)$ 
vanish in this limit (which we expect since in the commutative case 
$(\Bx\cdot\BL)$ is identically zero). 

Therefore to understand the structure 
of the new terms it is necessary to 
understand the properties of the operator 
$({\bf x\cdot L})$. 
Actually we can show that 
in the algebra ($\CA_N,\CL_N)$, $(\Bx\cdot\BL)$ does not vanish but 
is of the order $\alpha$.

Since 
\be 
[\BL_i,(\Bx\cdot \BL)]=0 \label{xLCommutesWithL}
\ee
holds, the action of $(\Bx\cdot \BL)$ on a polynomial of $\Bx_i$
is either a constant or proportional to the Casimir operator. An explicit 
calculation of the action on the highest weight vector shows that 
\be
(\Bx\cdot \BL)\Bx^n_{\pm}=\alpha{n(n+1)\over 2}\Bx^n_\pm\ .
\ee
Furthermore since
\bea
[(\BL\cdot\BL),\Bx_i]&=&2[\Bx_i-\BP_i]\ ,\\[0pt]
[(\Bx\cdot\BL),\Bx_i]&=&\alpha[\Bx_i-\BP_i]\ ,
\eea
$(\Bx\cdot\BL)$ has the same commutator with $\Bx_i$ as $(\BL\cdot\BL)$
and thus together with the relation (\ref{xLCommutesWithL}), 
on algebra level we can make the following identification 
\be
(\Bx\cdot\BL)={\alpha\over2}(\BL\cdot\BL)\ ,
\ee
which confirms our expectation above, i.e. $(\Bx\cdot\BL)$ 
vanishes in the commutative limit. 

Knowing the pair ($\gamma_\chi$, $\BD$) of chirality operator
and Dirac operator algebraically, we also introduce the spinors 
on which these operators are acting. In our formulation the 
Dirac spinor is simply $\CA_N\otimes\complex^2$. 
\be
\Psi=\mat{\psi^1\cr\psi^2}\ , \label{spinor}
\ee
where $\psi^a\in\CA_N$.

As usual the above spinor can be splitted into two sectors
according to the chirality by the projection operators
\be
\CP_\pm\equiv\half(1\pm\gamma_\chi)\ ,\label{ProjectionOperator}
\ee
as
\be
\Psi_\pm=\CP_\pm\Psi\ .
\ee
Correspondingly we can split the Dirac operator as 
\be
\BD=\BD_++\BD_-\ ,
\ee
where
\be
\BD_\pm={\ell\over\CN_N}\CP_\mp\{\Lambda+\id-{\alpha\over 2\ell^2}(\Bx\cdot\BL)\}\CP_\pm\ .\label{ChiralDirac}
\ee

\subsection{The Spectrum of the Dirac Operator}

To complete the discussion on the Dirac operator we compute its 
spectrum. For this end we first calculate the square of the
Dirac operator which is given by 
\be
\BD^2=-{1\over \ell^2} (\Sigma-\chi)^2\ .
\ee
After a tedious but straightforward calculation we obtain
\be
\BD^2= 
[(\BL\cdot\BL)+1+\Lambda]-{1\over \ell^2}(\Bx\cdot\BL)[(\Bx\cdot\BL)+\alpha(\Lambda+1)]\ ,
\ee
where we have used the following relation:
\bea
(\Sigma-\chi)(\Sigma-\chi)
&=&\Sigma\Sigma-\{\Sigma,\chi\}+\chi\chi\cr
&=&-\ell^2[(\BL\cdot\BL)+1+\Lambda]+(\Bx\cdot\BL)[\alpha+(\Bx\cdot\BL)
+\alpha\Lambda]\ .
\eea

Using the above formula, 
the spectrum of the Dirac operator can be represented by
using the Casimir operator as in the classical case.
For this purpose we introduce the 
"total angular momentum" as
\be
\BJ_i=\BL_i+\half\sigma_i \ .
\ee
The states are labeled by the eigenvalues of the Casimir operator 
as  
\be
\BJ^2\Psi_j=j(j+1)\Psi_j\ ,
\ee
where for the case of the algebra 
$\CA_N$, $0\leq j\leq {N+1\over 2}$ is an integer or 
a half integer depending on whether $N$ is odd or even, 
respectively.
Denoting the eigenvalues of $\BD^2$ by $\lambda^2_j$, i.e.,
\be
\BD^2\Psi_j=\lambda_j^2\Psi_j \ ,
\ee
their values are given by

\be
\lambda^2_j=(j+\half)^2
-{1\over N(N+2)}\big\{(j+\half)^4-(j+\half)^2\big\}\ . \label{eigensquare}
\ee

\proof
Using the relation $(\Bx\cdot\BL)={\alpha\over2}\BL^2$ we obtain
\be
\BD^2
=\BL^2+1+\Lambda-{\alpha^2\over4\ell^2}
\BL^2(\BL^2+2(\Lambda+1)) \ . \label{DiracSquare}
\ee
Then substituting the relations 
\bea
\Lambda&=&\BJ^2-\BL^2-{3\over4}\ ,\cr
\BJ^2&=&j(j+1)\ ,\cr
\BL^2&=&(j+s)(j+s+1)\ ,
\eea
into eq.(\ref{DiracSquare}), 
where $s=\pm\half$, 
we see that this equation depends only on $s^2={1\over4}$, and we obtain
the above result which depends only on $j$.
\qed 

The spectrum of the Dirac operator is then given by the square root 
of eq.(\ref{eigensquare}).  Taking the case where $\lambda_j>0$ as
\be
\BD\Psi_j=\lambda_j\Psi_j\ ,
\ee
since $\BD$ anticommutes with the chirality operator $\gamma_\chi$, we
obtain another spinor
\be
\BD(\gamma_\chi\Psi_j)=-\lambda_j(\gamma_\chi\Psi_j)\ ,
\ee
as expected.

It is interesting to compare the above results with the commutative case, 
i.e., to take the limit $N\rightarrow\infty$.
The above expression in eq.(\ref{eigensquare}) can be rewritten as 
\be
\lambda^2_j=(j+\half)^2[1
+{1-(j+\half)^2\over N(N+2)}]\ .
\ee
We obtain two different types of spectra. 

\begin{enumerate}
\item $N$ is even integer. Then $j$ is half integer and the spectrum is
\be
\lambda^2_j=1, \Big(4-{12\over N(N+2)}\Big),...,
\Big(n^2-{n^4-n^2\over N(N+2)}\Big),\cdots\ ,
\ee
where $1\leq n=j+\half\leq{N\over2}+1$.
\item $N$ is odd integer.  Then $j$ is integer and the spectrum is
\be
\lambda^2_j=\Big({1\over4}+{3\over 16N(N+2)}\Big),
\Big({9\over4}+{9\over 2N(N+2)}\Big),\cdots\ .
\ee
\end{enumerate}

For $N\rightarrow\infty$ the first case has 
a proper limit which gives the same spectrum as in the commutative case.
The appearance of the
second case can be understood rather easily.  
Following Berezin's general formulation, 
in principle any dimension of the Hilbert space $\CF_N$ can be chosen. 
However, when we consider the case where $N$ is odd, $\CF_N$ is the spin 
representation of $SU(2)$.
On the other hand, for the case 
where $N$ is even integer $\CF_N$ is 
a representation space of $SO(3)$.  
Since in the limit $N\rightarrow\infty$, 
with an appropriate reinterpretation of the measure, 
$\CF_N$ 
should approximate the Hilbert space of spherical harmonics, it must 
be a single valued representation. Therefore, the second case 
does not fit into our scheme and it is natural to 
restrict the quantization parameter $N$ to an even integer.

To conclude, although it seems that in the general formulation of 
the matrix algebra
representation of the fuzzy sphere any integer $N$ 
can be chosen for the $(N+1)\times (N+1)$ 
matrices, the above results shows that
$N$ must be an even integer in order to have a proper classical limit.

\bigskip

\subsection {Hilbert Space and New Algebra}
\medskip

In the previous section we constructed the Dirac operator by 
algebraic calculation, then we introduced the Dirac spinor as the 
space onto which the Dirac operator is acting in eq.(\ref{spinor}).
Below we shall see that also the structure of the 
algebra has to be reconsidered. 

Using the trace norm on the algebra the above spinor space 
becomes a Hilbert space with norm 
\be
\norm{\Psi}^2\equiv {1\over N+1}\sum_{a=1,2}\Tr{\psi^{a*}\psi^a} \ .
\ee
Thus we have the Hilbert space
\be
\hilbert_N \equiv \CA_N\otimes\complex^2\ .
\ee  
In order to complete 
Connes' K-cycle and thus put the noncommutative sphere into 
his framework, we have to reconsider the algebra of the 
fuzzy sphere. 
The reason for this is that 
the chirality operator defined above does not commute with the 
elements of ${\bf \CA}_N$. Instead we have 
\be
[\gamma_\chi,\BX_i]=0 \ ,
\ee
where 
\be
\BX_i=\Bx_i+{\alpha\over 2}\sigma_i\ .
\ee
Since the algebra should be trivial under the grading, 
we rather have to consider 
the algebra $\widetilde{\CA}_N$ of $\BX_i$ instead of the one of $\Bx_i$. 
The new generator $\BX_i$ satisfy the same commutation relation as
$\Bx_i$, i.e.,
\be
[\BX_i,\BX_j]=i\alpha \epsilon_{ijk}\BX_k \ .
\ee
Thus the new algebra $\widetilde{\CA}_N$ has the same multiplication rule
and we may consider it as a different realization of the algebra of the
fuzzy sphere.

In this way we have constructed the triple 
$(\widetilde{\CA}_N,\BD,\hilbert_N)$, starting from 
the algebra of the fuzzy sphere.  With this K-cycle we may construct 
the differential calculus 
and write down the field theory Lagrangian, which we will disscuss in a 
separate paper. In the remaining part of this paper, we want to 
examine a bit further the properties of the Dirac operator.

\bigskip

\subsection{Local Coordinates on Noncommutative Sphere}
\medskip

In ref.\cite{Jayewardena88}, the Schwinger model on $S^2$ is 
investigated and the classical Dirac operator on the sphere 
is discussed in great detail. In this context the
globally $SU(2)$ invariant form of the Dirac operator 
is compared with the Dirac operator in local 
coordinates. It is shown that the transformation from the covariant to 
the local frame is given 
as 
$\tilde{D}_{local}=u_cD_{covariant}u^{-1}_c$, where 
$u_c$ is a unitary matrix with determinat equal to unity.

It is interesting to examine the 
similar transformation for the Dirac operator on the noncommutative sphere.
This will help us to understand the meaning of a local coordinate patch in 
noncommutative geometry.

We define the matrix $u$ which gives the map discussed 
above by the requirement 
\be
\sigma_3 u  = u\gamma_\chi\label{conditionI}\ .
\ee
A lengthy but straightforward calculation yields 
\be
u={1 \over \sqrt{\alpha (N+1)}}\mat{\sqrt{\Bx_3+{\alpha\over 2}(N+2)}& 
{1\over \sqrt{(\Bx_3+{\alpha\over 2}(N+2))}}\Bx_-\cr 
 {-1\over \sqrt{(\Bx_3+{\alpha\over 2}N)}}\Bx_+ & 
\sqrt{\Bx_3+{\alpha\over 2}N}}\ ,
\ee
where $\Bx_{\pm}=\Bx_1\pm i\Bx_2$ and $\alpha$ is given in eq.(\ref{alpha}).

The above matrix becomes singular on the states satisfying
$(\Bx_3+{\alpha\over 2}N ) |v\rangle =0$ and 
$(\Bx_3+{\alpha\over 2}(N+2))|v'\rangle=0$.                              
Taking the basis given in eq.(\ref{Fockbasis}), 
the eigenvalues of the operator $\Bx_3$ are 
 limited by $\alpha{N\over 2}$, i.e., $\norm{\Bx_3}\leq\alpha{N\over 2}$, 
 thus we have to impose the following 
 restriction on the state in $\CF_N$
\be
(\Bx_3+{\alpha\over 2}N ) |v\rangle \not=0\ .\label{restrictions}
\ee
Therefore the following discussion holds only on the 
restricted Hilbert space, not on the whole space $\CF_N$. 

In the limit $N\rightarrow \infty$ which corresponds to the 
commutative case  and using eq.(\ref{alpha}) we obtain 
\be
u_{N\rightarrow\infty}={1\over \sqrt{2(1+{x_3\over\ell})}}
\mat{1+{x_3\over {\ell}}& 
{x_-\over\ell} \cr -{x_+\over\ell} &1+{x_3\over {\ell}}}\ .
\ee
Rewriting this expression in stereographic coordinates 
\be
x_1=2\ell^2z_1(\ell^2 +{\bf z}^2)^{-1}\ ,
\ee 
\be
x_2=2\ell^2z_2(\ell^2 +{\bf z}^2)^{-1}\ ,
\ee 
\be
x_3=\ell(\ell^2 -{\bf z}^2) (\ell^2 +{\bf z}^2)^{-1}\ ,
\ee 
where $z_i$ 
are the coordinates on the 2-sphere, we get
\be
u_{N\rightarrow\infty}(z)={1\over \sqrt{(\ell^2+z^2)}}
\mat{\ell& z_1-iz_2 \cr 
-(z_1+iz_2) &\ell}\ .
\ee
Up to a phase factor this is equivalent to the result 
obtained in \cite{Jayewardena88}.\footnote{The phase factor is 
\be
f={1\over \sqrt{2}}
\mat{(1+i)& 0 \cr 0 &(1-i)}\ .
\ee}

Under this transformation $u$ the coordinates $\BX_\pm$, $\BX_3$ 
transform into diagonal form as 
\be
u(\Bx_+ +{\alpha\over 2}\sigma_+)u^{\dagger}=
\mat{\sqrt{1+{\alpha\over \Bx_3+{\alpha\over 2}N}}\Bx_+  &0  \cr
0  &\sqrt{1-{\alpha\over \Bx_3 + {\alpha\over 2}N}}\Bx_+} \ ,
\ee
\be
u(\Bx_-+{\alpha\over 2}\sigma_-)u^{\dagger}=
\mat{\sqrt{1+{\alpha\over \Bx_3+{\alpha\over 2}(N+2)}}\Bx_-  &0  \cr
0  &\sqrt{1-{\alpha\over \Bx_3 + {\alpha\over 2}(N+2)}}\Bx_-} \ ,
\ee
and 
\be
u(\Bx_3 + {\alpha\over 2}\sigma_3)u^\dagger = \mat{\Bx_3+{\alpha\over 2}
& 0 \cr 0 & \Bx_3-{\alpha\over 2} }\ .
\ee
This is consistent since the chirality operator is now $\sigma_3$ and 
the elements which commute with the chirality operator are the diagonal
matrices. In this parametrization the Dirac operator has the form
\be
\BD'=\mat{0&\BD'_-\cr\BD'_+&0}\ ,
\ee
where $\BD'_\pm$ is defined by transforming the operators in 
eq.(\ref{ChiralDirac}).

As already mentioned, when we perform the above transformation
we have to restrict the states by eq.(\ref{restrictions}).
The interpretation of the restriction (\ref{restrictions}) on the 
states can be considered as an analogy to the idea of a local patch 
in the 
noncommutative geometry.

To complete the discussion, consider another transformation defined by 
\be
\sigma_3 u'  = -u'\gamma_\chi\ ,\label{conditionII}
\ee
which differs from condition (\ref{conditionI}) only by a sign. 
As a result we obtain a matrix $u'$
\be
u'={1 \over \sqrt{\alpha (N+1)}}\mat{\sqrt{\Bx_3-{\alpha\over 2}N}& 
{1\over \sqrt{(\Bx_3-{\alpha\over 2}N)}}\Bx_-\cr 
 {-1\over \sqrt{(\Bx_3-{\alpha\over 2}(N+2))}}\Bx_+ & 
\sqrt{\Bx_3-{\alpha\over 2}(N+2)}} \ .
\ee
This transformation matrix has also a singularity and thus correspondingly
we remove the state satisfying 
\be
 (\Bx_3-{\alpha\over 2}N)
|v'\rangle=0\ .\label{restrictions'}
\ee 
from the Hilbert space $\CF_N$.

Also in this case, 
 we obtain a chirality operator given
by $\sigma_3$ after the transformation 
and thus the algebra $\widetilde{\CA}$ is transformed 
into diagonal matrices and the Dirac operator has nonzero elements 
only in the off-diagonal components.

We may consider that the transformations $u$ and $u'$ are maps 
from symmetric coordinate to local coordinate systems.
The restriction on the states can be understood as the analogue of the 
definition of the regions of the local patches in the commutative case.

In the coherent state representation the restriction defines an 
open set on the sphere, since the complex value $v$ 
parametrizing the state $|v \rangle$ 
corresponds to a point on $S^2$. 
Consider the normalized state ${1\over(1+\overline{z}z)^{N/2}}|z\rangle$,
then the state which has to be removed corresponds to the point 
$z=\infty$, ($z=0$) for transformation $u$ ($u'$, respectively). 
Therefore, the transformation $u$ gives the "coordinate patch",
an open set of states over the noncommutative sphere where the "south pole" 
is excluded.  Correspondingly for $u'$ the "north pole" is excluded.

\bigskip

\section{Conclusions and Discussions}

In this paper we have defined the Dirac operator on the 
noncommutative sphere
and determined the triple $(\widetilde{\CA}_N, \BD, \hilbert_N)$
with grading operator $\gamma_\chi$, which defines Connes' 
K-cycle $(\BD,\hilbert_N)$ over the algebra $\widetilde{\CA}_N$.
We have started with the well known algebra of the fuzzy sphere $\CA_N$
and then have considered the bigger algebra ($\CA_N,\CL_N$)
 by including the derivations 
$\BL_i$ defined by the adjoint action of the coordinate function
$\Bx_i$.  Then, we defined the chirality operator and
the Dirac operator in the algebra $(\CA_N , \CL_N)\otimes M_2(\complex)$.
 The Dirac operator has been determined algebraically by requiring that 
it anticommutes with the chirality operator.  The construction is
performed purely algebraically. 

In order to determine the chirality operator
$\gamma_\chi$ we have required  
a ``correspondence principle'', i.e. the condition
that in the 
naive limit $N\rightarrow\infty$ this operator corresponds to the one 
of the commutative case. 
We also used this correspondence principle 
to choose the Dirac operator between two posibilities.
In this way we singled out a pair $(\BD,\gamma_\chi)$ in the algebra of
$(\CA_N,\CL_N)\otimes M_2(\complex)$. 

The spinors are introduced 
as the Hilbert space vectors on which these operators are acting, thus they
are vectors in the space $\CA_N\otimes \complex^2$.
Then, we have calculated the spectrum of the Dirac operator. 
As a result, we have found that there are two different sequences of the
quantized algebra   
depending on whether the quantization 
parameter $N$ is an ever or an odd integer, i.e., 
$\{\CA_{2k}\}$ and $\{\CA_{2k+1}\}$ with integer $k$. 
We have seen that a restriction of 
$N$ to even integers gives the desired classical 
limit. This requirment means that the Hilbert space $\CF_N$ appearing in 
Berezin's quantization has to be a single valued representation 
of the rotation group of $S^2$.

We also defined the Connes' triple $(\widetilde{\CA}_N,\BD,\hilbert_N)$. 
The algebra of the noncommutative sphere $\widetilde{\CA}_N$ is obtained
by modifying the original algebra $\CA_N$. The modification is required
since the algebra $\CA_N$ does not commute with the chirality operator and 
the new algebra $\widetilde{\CA}_N$ is defined 
as a subalgebra of $\CA_N\otimes M_2(\complex)$ by the requirment that
it commutes with the chirality operator.

Finally we considered the transformation of the chirality 
operator $\gamma_\chi$
into the standard form, i.e., $\sigma_3=\mat{1&0\cr0&-1}$. We found two 
unitary matrices $u$ and $u'$ defined 
by eqs.(\ref{conditionI}) and (\ref{conditionII}). 
However the unitary matrices $u$ and $u'$ are not defined on 
the whole set of
states in the Hilbert space $\CF_N$ since they have singularities 
depending on the eigenvalues of the operator $\Bx_3$.  
We meet the similar situation in the commutative case when we transform the
coordinate system from the symmetric coordinates to the local ones. 
To avoid these singularities, 
in analogy to the commutative case we restrict the vectors in 
the Hilbert space
$\CF_N$.  In the coherent state representation, this restriction 
corresponds 
to considering the states labeled by the points in a certain 
open set on $S^2$. 
Actually, under the maps given by $u$ and $u'$ the algebra 
$\tilde{A}_N$ is diagonalized, and the set of coherent states 
given by the open set of parameters restricted to the region around 
the north pole and south pole, respectively, can be interpreted as the 
analogue of a local patch in the noncommutative case. 

The starting point of our construction 
is near to the original idea of ref.\cite{GrosMado92} 
where the fermion fields are also considered as $\CA\otimes\complex^2$ with 
 the association to the classical case in ref.\cite{Jayewardena88}. 
However unlike in the 
papers by \cite{GrosKlim95}
we do not make use of the supersymmetry 
algebra in order to define the Dirac operator. 
As a consequence our results are 
different from their Dirac operator.

Especially we have found 
 that in the noncommutative case there is a contribution 
from an operator $(\Bx\cdot\BL)$ which is zero in the commutative limit. 
Without this contribution of $(\Bx\cdot\BL)$, in the noncommutative 
space the Dirac operator does not anticommute 
with the chirality operator.
As a result, the spectrum of the Dirac operator has a correction
term compared to the commutative case.

With the results obtained here we can 
construct the differential calculus \`a la Connes. 
This is now in preparation. 

Finally we want make a comment on the property of the Planck constant 
in our formulation. As we mentioned, in the method described here 
the formulation of the function algebra 
is independent of the dimension of the Hilbert space. 
The $\BN$ and thus the Planck constant $\alpha$
appears as an operator in the algebra. 
The dependence on the
dimension of the Hilbert space enters when computing the 
expectation value with respect of a certain coherent state or the trace 
in the Hilbert space.

\bigskip

{\large{\bf Acknowledgement}}

The authors benefited from discussions with 
M. Bordemann, O. Grandujan and M. Pillin. S.W. would like to thank
 K. Osterwalder for his hospitality during the stay in ETH where
 this work began. He also thanks the Canon Foundation in Europe for 
 supporting that stay. 
 U.C. would like to acknowledge the Japan Society for Promotion of Science for 
financial support.

\end{document}